\documentclass[sn-mathphys]{sn-jnl}

\jyear{2022}%

\theoremstyle{thmstyleone}%
\theoremstyle{thmstyletwo}%

\theoremstyle{thmstylethree}%

\begin{document}

\title[Comments on: Hybrid Semiparametric Bayesian Networks]
  {Comments on: \linebreak Hybrid Semiparametric Bayesian Networks}

\author*[1]{\fnm{Marco} \sur{Scutari}}\email{scutari@bnlearn.com}

\affil*[1]{
  \orgname{Istituto Dalle Molle di Studi sull'Intelligenza Artificiale (IDSIA)},
  \orgaddress{\city{Lugano}, \country{Switzerland}}%
}

\maketitle

This is an interesting paper that distils structure learning in Bayesian
networks (BNs) and kernel methods in a quest to produce more flexible
distributional assumptions. Conditional (linear) Gaussian Bayesian networks
(CGBNs) have been well explored in the literature for some time, to the point
that they now appear in many recent textbooks \cite{koller,crc21,kjaerluff}. The
authors address one of the key limitations of CGBNS, that they can only capture
linear dependencies between the continuous variables they contain, and remove it
by replacing (mixtures of) linear regression models with more general kernel
densities. Dependencies between discrete variables were already flexible,
because the conditional probability tables that parametrise them essentially act
as a saturated model \cite{cpts}. It is not obvious that more flexibility will
produce better models for whatever task we have in mind: it can also lead to
overfitting, instability and hyperparameter tuning problems. However, the
accuracy of reconstruction demonstrated by the proposed Hybrid Semiparametric
BNs (HSPBNs) is encouraging.

Like all good papers, it raises interesting questions in its construction.

\emph{How to measure structural distances?} Common distributional assumptions in
BNs, including CGBNs, assign the same type of distribution to a node in all
possible network structures. Therefore, the presence of an arc denotes the same
general type of dependence in all possible structures as well. However, this is
no longer the case in HSPBNs because continuous nodes can have either parametric
or nonparametric characterisations for the same arcs. The authors acknowledge
this by complementing the Structural Hamming Distance \cite[SHD;][]{mmhc} with a
Type Hamming Distance (THD) in their experimental evaluation. Should we combine
them in a single measure by adding colours to the arcs and extending SHD to
count different colours as errors? And how should we weight such errors compared
to false positive and false negative arcs? Then there is also the question of
whether we should update the definition of equivalence classes: it is used in
constructing SHD and it has wide implications in our interpretation of BNs. In a
CGBNs, continuous variables are assumed to be jointly distributed as a
multivariate normal: that ensures that arcs that are not compelled can be
oriented in either direction while producing networks in the same equivalence
class. It is not obvious that this is the case with nonparametric nodes. We
would also assume that any single node must have the same distribution in all
the BNs in the same equivalence class, which means extending its
characterisation as well to consider arc colours.

\emph{How to implement constraint-based learning with kernel-based nonparametric
local distributions?} Since the heuristic algorithms used in constraint-based
learning are distribution-agnostic, the question becomes how to define a
suitable conditional independence test. The answer is not trivial because we
cannot easily construct likelihood-ratio tests from the likelihoods used in the
cross-validated score $S^k_\mathrm{CV}(\mathcal{D}, \mathcal{G})$. Firstly,
those likelihoods are only defined for continuous variables conditional on
discrete ones which is problematic when testing the independence of discrete
variables conditional on continuous ones. Secondly, it is unclear what the
degrees of freedom of the test would be: computing effective degrees of freedom
from the kernel transform is possible \cite{elemstatlearn}, but not obviously
appropriate. One option is to look for inspiration at those existing kernel
tests that have been extended to test conditional independence like the
Hilbert-Schmidt independence criterion \cite[HSIC;][]{hsic, chsic}.

A second option would be to look at BN structure learning approaches such as
\cite{cussens} that use kernels for both continuous and discrete variables. The
resulting flexibility in defining the nature of conditional independence
relationships in the BN could also address the remaining limitation of CGBNs
that is still present in HSPBNs: \emph{removing the constraint that continuous
nodes cannot be parents of discrete nodes}. The impact of this limitation cannot
be overstated: it prevents CGBNs and HSPBNs from being used as causal models in
the general case because the direction of arcs connecting discrete and
continuous variables is fixed and has nothing to do with the cause-effect
relationship present in the data we learn such BNs from. In turn, the directions
of adjacent arcs may not necessarily reflect cause-effect relationships either
because of the cascading effects of incorrect arc inclusions in structure
learning.

\backmatter



\begin{thebibliography}{9}
\ifx \bisbn   \undefined \def \bisbn  #1{ISBN #1}\fi
\ifx \binits  \undefined \def \binits#1{#1}\fi
\ifx \bauthor  \undefined \def \bauthor#1{#1}\fi
\ifx \batitle  \undefined \def \batitle#1{#1}\fi
\ifx \bjtitle  \undefined \def \bjtitle#1{#1}\fi
\ifx \bvolume  \undefined \def \bvolume#1{\textbf{#1}}\fi
\ifx \byear  \undefined \def \byear#1{#1}\fi
\ifx \bissue  \undefined \def \bissue#1{#1}\fi
\ifx \bfpage  \undefined \def \bfpage#1{#1}\fi
\ifx \blpage  \undefined \def \blpage #1{#1}\fi
\ifx \burl  \undefined \def \burl#1{\textsf{#1}}\fi
\ifx \doiurl  \undefined \def \doiurl#1{\url{https://doi.org/#1}}\fi
\ifx \betal  \undefined \def \betal{\textit{et al.}}\fi
\ifx \binstitute  \undefined \def \binstitute#1{#1}\fi
\ifx \binstitutionaled  \undefined \def \binstitutionaled#1{#1}\fi
\ifx \bctitle  \undefined \def \bctitle#1{#1}\fi
\ifx \beditor  \undefined \def \beditor#1{#1}\fi
\ifx \bpublisher  \undefined \def \bpublisher#1{#1}\fi
\ifx \bbtitle  \undefined \def \bbtitle#1{#1}\fi
\ifx \bedition  \undefined \def \bedition#1{#1}\fi
\ifx \bseriesno  \undefined \def \bseriesno#1{#1}\fi
\ifx \blocation  \undefined \def \blocation#1{#1}\fi
\ifx \bsertitle  \undefined \def \bsertitle#1{#1}\fi
\ifx \bsnm \undefined \def \bsnm#1{#1}\fi
\ifx \bsuffix \undefined \def \bsuffix#1{#1}\fi
\ifx \bparticle \undefined \def \bparticle#1{#1}\fi
\ifx \barticle \undefined \def \barticle#1{#1}\fi
\bibcommenthead
\ifx \bconfdate \undefined \def \bconfdate #1{#1}\fi
\ifx \botherref \undefined \def \botherref #1{#1}\fi
\ifx \url \undefined \def \url#1{\textsf{#1}}\fi
\ifx \bchapter \undefined \def \bchapter#1{#1}\fi
\ifx \bbook \undefined \def \bbook#1{#1}\fi
\ifx \bcomment \undefined \def \bcomment#1{#1}\fi
\ifx \oauthor \undefined \def \oauthor#1{#1}\fi
\ifx \citeauthoryear \undefined \def \citeauthoryear#1{#1}\fi
\ifx \endbibitem  \undefined \def \endbibitem {}\fi
\ifx \bconflocation  \undefined \def \bconflocation#1{#1}\fi
\ifx \arxivurl  \undefined \def \arxivurl#1{\textsf{#1}}\fi
\csname PreBibitemsHook\endcsname

\bibitem{koller}
\begin{bbook}
\bauthor{\bsnm{Koller}, \binits{D.}},
\bauthor{\bsnm{Friedman}, \binits{N.}}:
\bbtitle{Probabilistic Graphical Models: Principles and Techniques}.
\bpublisher{MIT Press},
\blocation{Cambridge, MA, USA}
(\byear{2009})
\end{bbook}
\endbibitem

\bibitem{crc21}
\begin{bbook}
\bauthor{\bsnm{Scutari}, \binits{M.}},
\bauthor{\bsnm{Denis}, \binits{J.-B.}}:
\bbtitle{Bayesian Networks with Examples in R},
\bedition{2nd} edn.
\bpublisher{Chapman \& Hall},
\blocation{Boca Raton, FL, USA}
(\byear{2021})
\end{bbook}
\endbibitem

\bibitem{kjaerluff}
\begin{bbook}
\bauthor{\bsnm{Kj{\ae}rluff}, \binits{U.B.}},
\bauthor{\bsnm{Madsen}, \binits{A.L.}}:
\bbtitle{Bayesian Networks and Influence Diagrams: A Guide to Construction and
  Analysis},
\bedition{2nd} edn.
\bpublisher{Springer},
\blocation{New York, USA}
(\byear{2013})
\end{bbook}
\endbibitem

\bibitem{cpts}
\begin{barticle}
\bauthor{\bsnm{Rijmen}, \binits{F.}}:
\batitle{{Bayesian Networks with a Logistic Regression Model for the
  Conditional Probabilities}}.
\bjtitle{International Journal of Approximate Reasoning}
\bvolume{48}(\bissue{2}),
\bfpage{659}--\blpage{666}
(\byear{2008})
\end{barticle}
\endbibitem

\bibitem{mmhc}
\begin{barticle}
\bauthor{\bsnm{Tsamardinos}, \binits{I.}},
\bauthor{\bsnm{Brown}, \binits{L.E.}},
\bauthor{\bsnm{Aliferis}, \binits{C.F.}}:
\batitle{{The Max-Min Hill-Climbing Bayesian Network Structure Learning
  Algorithm}}.
\bjtitle{Machine Learning}
\bvolume{65}(\bissue{1}),
\bfpage{31}--\blpage{78}
(\byear{2006})
\end{barticle}
\endbibitem

\bibitem{elemstatlearn}
\begin{bbook}
\bauthor{\bsnm{Hastie}, \binits{T.}},
\bauthor{\bsnm{Tibshirani}, \binits{R.}},
\bauthor{\bsnm{Friedman}, \binits{J.}}:
\bbtitle{The Elements of Statistical Learning: Data Mining, Inference, and
  Prediction},
\bedition{2nd} edn.
\bpublisher{Springer}, \blocation{???}
(\byear{2009})
\end{bbook}
\endbibitem

\bibitem{hsic}
\begin{bchapter}
\bauthor{\bsnm{Gretton}, \binits{A.}},
\bauthor{\bsnm{Fukumizu}, \binits{K.}},
\bauthor{\bsnm{Teo}, \binits{C.-H.}},
\bauthor{\bsnm{Song}, \binits{L.}},
\bauthor{\bsnm{Sch{\"o}lkopf}, \binits{B.}},
\bauthor{\bsnm{Smola}, \binits{A.J.}}:
\bctitle{{A Kernel Statistical Test of Independence}}.
In: \bbtitle{Advances in Neural Information Processing Systems},
pp. \bfpage{585}--\blpage{592}
(\byear{2008})
\end{bchapter}
\endbibitem

\bibitem{chsic}
\begin{bchapter}
\bauthor{\bsnm{Doran}, \binits{G.}},
\bauthor{\bsnm{Muandet}, \binits{K.}},
\bauthor{\bsnm{Zhang}, \binits{K.}},
\bauthor{\bsnm{Sch{\"o}lkopf}, \binits{B.}}:
\bctitle{{Permutation-Based Kernel Conditional Independence Test}}.
In: \bbtitle{Uncertainty in Artificial Intelligence},
pp. \bfpage{132}--\blpage{141}
(\byear{2014})
\end{bchapter}
\endbibitem

\bibitem{cussens}
\begin{barticle}
\bauthor{\bsnm{Handhayani}, \binits{T.}},
\bauthor{\bsnm{Cussens}, \binits{J.}}:
\batitle{{Kernel-based Approach for Learning Causal Graphs from Mixed Data}}.
\bjtitle{Proceedings of Machine Learning Research}
\bvolume{138},
\bfpage{221}--\blpage{232}
(\byear{2020})
\end{barticle}
\endbibitem

\end{thebibliography}
\end{document}